\documentclass{aa}  
\usepackage{natbib}
\bibpunct{(}{)}{;}{a}{}{,}
\usepackage{graphicx}
\usepackage{epstopdf}
\usepackage{stfloats}
\usepackage{float}
\usepackage{txfonts}
\usepackage{caption}
\usepackage[colorlinks,
            linkcolor=red,
            anchorcolor=blue,
            citecolor=blue]{hyperref}
 
\usepackage{amstext}

\begin{document} 

   \title{Formation of hot subdwarf B stars with neutron star components}
   \titlerunning{Formation of sdB with NS components}

   \author{You Wu
          \inst{1,2,3},
          Xuefei Chen\inst{1,2,4},
          Zhenwei Li\inst{1,2,3}
          \and
          Zhanwen Han\inst{1,2,4}
          }
   \authorrunning{Y. Wu
          \inst{1,2,3},
          X. Chen\inst{1,2,4},
          Z. Li\inst{1,2,3}
          \and
          Z. Han\inst{1,2,4}
          }
   \institute{
   $^1$Yunnan Observatories, Chinese Academy of Sciences,
            Kunming 650216, China; youwu@ynao.ac.cn, cxf@ynao.ac.cn\\
           $^2$ Key Laboratory for the Structure and Evolution of Celestial Objects, Chinese Academy of Sciences, Kunming 650216, China\\
            $^3$ University of the Chinese Academy of Science, Beijing 100049, China\\
            $^4$ Center for Astronomical Mega-Science, Chinese Academy of Sciences, 20A Datun Road, Chaoyang District, Beijing, 100012, China\\
             }

   \date{}

 
  \abstract
   {Binary population synthesis predicts the existence of subdwarf B stars (sdBs) with neutron star (NS)  or black hole (BH) companions. Several works have been dedicated to finding such systems,  but none has been confirmed yet. Theoretically, the formation of sdBs with white dwarf (WD) and main sequence (MS) companions has been well investigated, while those with NS or BH companions remain to be explored further. }
   {We systematically investigate the formation of sdB+NS binaries from binary evolution and aim to obtain some clues for a search for such systems. }
   {We started from a series of MS+NS systems and determined the parameter spaces for producing sdB+NS binaries from the stable Roche-lobe overflow (RLOF) channel and from the common envelope (CE) ejection channel. The  parameters for sdB+NS binaries were obtained from detailed binary evolution calculation with the code called \emph{\textup{modules for experiments in stellar astrophysics}} (MESA), and the CE parameters were given by the standard energy budget for CE evolution. The MS star had an initial mass ranging from 0.8 to $5 \,M_\odot$.  Various NS accretion efficiencies and NS masses were examined to investigate the effects they have. We show the characteristics of the produced sdB+NS systems, such as the mass of components, orbital period, the semi-amplitude of the radial velocity ($K$), and the spin of the NS component.}
 {sdB+NS binaries can be produced either from stable RLOF or from CE ejection. In the stable RLOF channel, sdBs can be formed when the donor starts mass transfer close to the tip of the giant branch if the donor has an initial mass $\leq 2.0M_{\odot}$. For more massive donors, sdBs can be formed when the donor starts mass transfer during the Hertzsprung gap or near the end of the MS. The orbital period of sdB+NS binaries produced in this way ranges from several days to more than 1000 days and moves toward the short-period ($\sim$ hr) side with increasing initial MS mass. The highest $K$ is about $150 {\rm km\ s^{-1}}$ for an MS star of initially  $5M_\odot$ . However, the sdB+NS systems that result from CE ejection have very short orbital periods and then high values of $K$ (up to $800 \rm km\ s^{-1}$). Such systems are born in very young populations (younger than 0.3 Gyr) and are potential gravitational wave sources that might be resolved by the Laser Interferometer Space Antenna (LISA) in the future. Gravitational wave radiation may again bring them into contact on a timescale of only $\sim$ Myr. As a consequence, they are rare and hard to discover. The pulsar signal is likely a feature of sdB+NS systems caused by stable RLOF, and some NS components in sdB binaries may be millisecond pulsars. Various NS accretion efficiencies and NS masses change some properties of sdB+NS binaries, but not our general results.}
   {}


     \maketitle
%

\section{Introduction}

Hot subdwarf-B stars (sdBs) are located at the extreme horizontal branch (EHB) in a Hertzsprung-Russell diagram (HRD). These objects have relatively high temperatures ($\mathit{T}_{\mathrm{eff}}\approx 20000- 40000\mathrm{K}$) and high surface gravities ($\log \mathit{g}$ between $\sim$ 5.0 and 6.5). The structure of an sdB typically consists of a helium-burning core and a thin hydrogen-rich envelope ($\lesssim 0.02 \,M_\odot$) \citep{Heber1986}. Theoretical studies show a wide mass range for sdB stars from 0.3 to 0.8$M_\odot$ in \citet{Han2002,Han2003} and from 0.35 to 1.7$M_\odot$ in the recent study of \citet{2018arXiv180203018G}.
   
SdBs are in a special stage in stellar evolution, which occurs after a star has lost almost the whole hydrogen envelope before it ignites helium degenerately or non-degenerately in the core. It has been discovered that two types of pulsations exist in some sdBs: pressure-mode (p-mode) pulsations \citep{Charpinet}, and gravity-mode (g-mode) pulsations \citep{Fontaine}, but both may also be present in an sdB. An sdB stays on the EHB for roughly $10^8$ years and directly evolves along with the white dwarf (WD) cooling track after its core helium has been exhausted. \citet{Heber2016} provided a comprehensive review of sdBs as a whole.

It is widely accepted that the sdBs are formed from binary evolution, that is, common envelope (CE) ejection, stable Roche-lobe overflow (RLOF), and the merger of double helium WD (He-WD) {\citep{2018MNRAS.476.5303S}.} sdB binaries with short orbital periods account for a comparatively large portion of sdBs, and their companions are generally main-sequence (MS) stars or WDs \citep{Maxted,Napiwotzki2004}. The sdBs with massive WD companions, such as KPD 1930+2752 and CD-30 11223, have been considered good candidates of Type Ia supernovae (SN Ia) progenitors \citep{Maxted2000,Geier2007,2013ASPC..469..373G,Wang2009,Vennes2012}. Numerous investigations have been reported that focused on the formation of sdB+WD and sdB+MS systems \citep{Han2002,Han2003,Chen2013}. The formations of sdB+neutron star (NS) and sdB+black hole (BH) systems, however, have so far not been studied systematically and remain to be explored. 

In recent years, considerable attentions have been paid to the systems of sdBs coupled with a massive companion on observations, 
e.g. the project Massive Unseen Companions to Hot Faint Under-luminous Stars from SDSS (MUCHFUSS). 
This project seeks to find sdBs with massive compact companions such as massive WDs (> $1.0 \,M_\odot$), NSs or stellar mass BHs \citep{Geier2011}. 
In this project, objects which have high radial velocity (RV) variations from Sloan Digital Sky Survey (SDSS) have been selected as good candidates and re-observered to obtain the medium resolution spectra. 
Until now, there are 129 sdB samples discovered by MUCHFUSS \citep{Geier2015,Geier2017a}. The majority of the sdB samples have RV semi-amplitude $\mathit{(K)}$ values in a range of 50 - 160 $\mathrm{ km\ s^{-1}}$, and the highest $\mathit{K}$ value is $\mathrm{359 km\ s^{-1}}$. Most objects exceeding $\mathit{d} \sim$ 3 kpc may be located in the Galactic halo due to the fact that the coverage of SDSS roughly belongs to high Galactic latitudes. No sdBs with NS or BH companions have been confirmed yet and the fraction of close sdB+NS/BH binaries has been constrained to be less than 1.5\% \citep{Geier2017a} based on the results from the MUCHFUSS project. 

A possible candidate for sdB+NS binaries is the millisecond pulsar, PSR J1816+4510, which has been identified by \citet{Kaplan2013}. Its companion has atmospheric parameters similar to those of an sdB, that is, $\mathit{T}_{\mathrm{eff}}\approx 16000\mathrm{K}$, $\log \mathit{g}\approx 4.9$. However, \citet{2014A&A...571L...3I} suggested that the companion might be an extremely low-mass proto-He WD. The object HD 49798 is a subdwarf O star (sdO) \citep{Jaschek1963} with a massive compact companion ($1.28\pm 0.05M_\odot$). It is the first sdO star to have been detected with X-ray emission, and the companion spins at 13.2s in a 1.55day orbit \citep{Thackeray1970,Israel1996}. Some hypotheses have been made on the type of the compact companion. Most recently, the companion has been suggested to be an NS based on the spin-up rate derived from the observations of X-ray pulsations. The possibility of a massive WD companion cannot be excluded, however \citep{Mereghetti2016}. Owing to the lack of direct evidence for sdB+NS/BH binaries from observations, it is doubtful whether such systems might indeed be produced, since binaries are likely to be disrupted by supernova explosions when the NS/BH forms. However, the existence of numerous X-ray binaries, the majority of which have NS/BH companions, removes this concern. Meanwhile, \citet{2011MNRAS.416.2130T} modeled the formation of the massive millisecond pulsar binary PSR J1614-2230, and clearly showed the progenitor phase for a NS+sdB binary. 

NSs can be produced by core-collapse supernovae (CC), electron-capture supernovae (ECSNe), and accretion-induced collapse (AIC) from an oxygen-neon-magnesium (ONeMg) WD \citep{Miyaji1980,Nomoto1984,Nomoto1987,1987Natur.329..310M}. NS with the lowest ( $ 1.15M_{\odot }\sim 1.22M_{\odot }$) and highest ($\ge \sim 1.4M_{\odot }$) mass are both expected to be produced by the CC scenario, while NSs with masses of about 1.25$M_{\odot }$ are more likely from the ECSNe and the AIC scenario \citep{1996ApJ...457..834T,2011BASI...39....1V}. The natal kicks of NSs imparted by ECSNe and by AIC are believed to be much smaller than the kick imparted in the CC scenario \citep{Pfahl2002}. \citet{Zhu2015} studied the formation of millisecond pulsars (MSPs) through a population synthesis approach and showed that approximately 58\% of radio MSPs are produced in the CC, 35\% in the ECSNe, and 7\% in the AIC scenario.
Taking into account the asymmetric kicks contributed by the formation of NS/BH in the population synthesis model, \citet{Nelemans2013} predicted that about 1\% of the sdB binaries have NS companions and about 0.1\% have BH companions. However, the details of the formation process of sdB+NS binaries have not been considered adequately in that study.

The primary objective of this research is to systematically study the formation of sdB+NS systems. By associating simulation results with the observations, we could obtain some valuable hints for observations and also boost the insights into the evolutionary process of sdBs. The remainder of this paper is structured as follows. Sect. 2 introduces the models, simulating methods, and assumptions we made for mass transfer process and NS accretion. The results are presented in Sect. 3 for the stable RLOF and in Sect. 4 for the CE ejection channel. Our main conclusions are summarized in Sect. 5, followed with a discussion of future research. 
 
\section{Models}

We started from binaries consisting of an MS star and an NS companion, 
and ignored the evolution before the formation of NSs because of the uncertainties in supernova explosions. 
Based on the binary model for the sdB formation \citep{Han2002,Han2003}, sdB+NS binaries can be produced through either the CE ejection or the stable RLOF channel. 
In our study, the NS was set to be of canonical mass, $1.4 \,M_\odot$ , and was treated as a point mass, similar to \citet{Lin2011}.  
The masses of the MS star ranged from $0.8 \,M_\odot$ to $5.0 \,M_\odot$ 
in steps of $\Delta \log \left ( M/M_{\odot} \right )= 0.1$, that is, $M_{\rm d}^{\rm i}=0.8 \,M_\odot$, $1.0 \,M_\odot$, $1.26 \,M_\odot$, $1.6 \,M_\odot$, $2.0 \,M_\odot$, 
$2.5 \,M_\odot$, $3.2 \,M_\odot$, $4.0 \,M_\odot$ , and $5.0 \,M_\odot$. 
Systems with MS stars more massive than $5.0 \,M_\odot$ were not considered here since the RLOF between an NS and such a high-mass MS star is probably dynamically unstable and therefore leads to a CE.\footnote{ However, MS stars with mass $ > 5.0 \,M_\odot$ have a strong wind, and NSs can accrete mass from the MS stars via stellar wind, resulting in the formation of high-mass X-ray binaries \citep{Tauris2017}.}

Whether RLOF is dynamically stable or unstable depends on the evolutionary phase of the initial MS star (the donor) and the mass ratio at the onset of RLOF. In general, the critical mass ratio $q_{\rm c}$ for dynamically stable RLOF is considered to be around 4.0 for the donors on MS or the\ Hertzsprung gap (HG). The main argument comes from the case that the donor starts RLOF on the giant branch (GB). As is well known, if the mass transfer is conservative, the value of $q_{\rm c}$ is $\sim 2/3$ for a fully convective donor, predicted from a polytrope with a polytropic index of 1.5. The assumption of non-conservative mass transfer process and the existence of the core in a giant star cause a higher value of $q_{\rm c}$ , but not exceeding$\sim 1.0$ by much. However, \citet{2000ApJ...530L..93T} showed that donor stars up to $ 6.0 \,M_\odot$ can result in stable RLOF in a $1.3 \,M_\odot$ NS accretor if the convective envelope of the donor star is not too deep, clearly demonstrating that the results from polytropic models need to be improved.\footnote{ Fig.1 in \citet{2000ApJ...530L..93T} shows that an NS+sdB binary is produced as a system evolves from stage $f$ to $g$.} The study of detailed binary evolution calculations showed that the value of $q_{\rm c}$ can be as high as $\sim 2.0$ for giant donors \citep{2008MNRAS.387.1416C}. \citet{Pavlovskii2014} also showed $q_{\rm c}$ to be about 1.7-2.3 for conservative mass transfer from giant donors. The very recent study from Ge et al. (in preparation) shows that $q_{\rm c}$ changes with donor mass and donor core mass on the GB, and the values of $q_{\rm c}$ are approximately equal to 1.5-3.0 in their study.
Therefore, the mass transfer is always dynamically stable in our models if the donor starts RLOF on the MS or during HG. When the donor starts mass transfer on the GB, however, the mass transfer is artificially assumed to be dynamically unstable for $M_{\textrm{d}}^{\textrm{i}}\geq 3.2 \,M_\odot$.\footnote{ We have examined some cases with $3.2M_{\odot}$ giant donors, and found that the mass transfer rate increases dramatically to exceed $1.0M_{\odot}$/yr after it reaches $10^{-4} M_{\odot}/yr$. Then the code stalled due to the very short and unreasonable timescales. The dynamical instability cannot be avoided at such high mass transfer rates. }

For the systems undergoing stable RLOF, the donor star loses most of its envelope during RLOF, then an sdB forms if the helium core is ignited after RLOF. 
We used the binary module of code called modules for experiments in stellar astrophysics (MESA, version 7684) 
to simulate the evolution of the models. 
The MESA code has increased in functionality and added a series of ameliorations 
for mass transfer process, 
and it has also somehwat improved on the He-core flash during stellar evolution \citep{Paxton2011,Paxton,Paxtona}. 
We studied Population I stars (with a metallicity of $Z=0.02$). The general parameters were adopted from the calibration of the Sun: the mixing length parameter $\alpha_{\rm MLT}=1.91076$ and extra overshooting control $f_{\rm ov}\approx 0.0029$, in units of pressure scale height for an exponentially overshooting model. 

The mass transfer rate was calculated from the scheme of \citet{Ritter1988}, and the maximum value was artificially set to be $10^{-4} M_\odot {\rm yr}^{-1}$. Similar to  \citet{Lin2011}, the accretion rate of the NS was limited by Eddington limit, and 90\% ($\varepsilon=0.9$) of the transferred mass remained on the NS surface if the mass transfer rate meets sub-Eddington mass transfer rates. The other 10\% of the transferred matter was ejected from the NS and took away the specific angular momentum as that contained by the NS. To investigate the influence of the accretion efficiency and NS mass, we also examined the cases for $\varepsilon =0.5, 0.7$ and an initial NS mass $M_{\rm NS}^{\rm i}=1.25 \,M_\odot$. 

For each of the donors, we first determined the minimum initial orbital period, $P_{\rm min}^{\rm i}$,  with which 
the donor retains the minimum He core for He ignition after the end of RLOF. In other words, the mass of He core cannot grow up to the minimum core mass for He ignition during the mass transfer if the initial orbital period $P^{\rm i}<P_{\rm min}^{\rm i}$. The maximum initial period $P_{\rm max}^{\rm i}$ is defined as the period beyond which the He core of the donor is ignited before the end of mass transfer and too much of the H-rich envelope remains on the surface, leaving a red clump star rather than an sdB \citep{Girardi2016}. We increased $P^{\rm i}$ from $P_{\rm min}^{\rm i}$ at regular intervals to determine $P_{\rm max}^{\rm i}$ and obtained the parameter space for the formation of sdB+NS binaries. Finally, we list the evolutionary consequences of our models, such as the component masses $M_{\rm sdB}$ and $M_{\rm NS}$, final orbital periods $P_{\rm f}$,  the RV semi-amplitude $K$, and the spin period of NS $P_{\rm{NS}}$. 

If the mass transfer is dynamically unstable, a CE forms soon after mass transfer starts. A reasonable and commonly used approximation for the CE evolution is called standard energy budget \citep{Webbink1984,Livio1988,DeKool1990}, that is, the CE can be ejected if 
\begin{equation}
   \alpha _{\textrm{CE}} \Delta E_{\textrm{orb}} \geqslant E_{\rm gr} +\alpha_{\rm th} E_{\rm th} 
   ,\end{equation}
where $\Delta E_{\textrm{orb}}$ is the orbital energy released during the spiraling process \citep{Iben1984},
$E_{\textrm{gr}}$ is the gravitation binding energy, and $E_{\textrm{th}}$ is the internal energy  \citep{Han1994}, involving terms of ionization energy, the basic thermal energy for a simple perfect gas energy of radiation, and the Fermi energy of a degenerate electron gas.
Parameters of $ \alpha _{\textrm{CE}}$ and $ \alpha _{\textrm{th}}$ are the efficiencies of $\Delta E_{\rm orb}$ and $E_{\rm th}$, respectively, indicating the fraction of the energy to be used to eject the envelope. For given $ \alpha _{\textrm{CE}}$ and $ \alpha _{\textrm{th}}$, we can determine whether the envelope of the donor can be successfully ejected, and obtain the final orbital period of the system if the envelope can be ejected.
We here set the upper limits for the two parameters, that is, $ \alpha _{\textrm{CE}}=\alpha _{\textrm{th}}=1$, which give the maximum parameter space for CE ejection. To obtain the values of $E_{\textrm{gr}}$ and $E_{\textrm{th}}$, we first evolved the donors to the onset of CE with MESA, then integrated the donors from the base of the envelope to the surface following \citet{Han1994}.

For systems undergoing unstable RLOF, we determined the minimum core mass for He ignition for each donor, $M^{\rm min}_{\rm c}$, and whether the CE could be ejected at that point, as described above. If the CE could be ejected, an sdB+NS binary is produced. Denoting $P_{\rm ig}$ and $P_{\textrm{tip}}$ as the orbital periods with which the donor starts CE at $M^{\rm min}_{\textrm{c}}$ and at the tip of giant branch, respectively, the orbital period range for producing sdB+NS binaries from the CE ejection channel is between $P_{\rm ig}$ and $P_{\textrm{tip}}$. If the CE could not be ejected at $M^{\rm min}_{\textrm{c}}$, we need to further determine the minimum initial orbital, $P_{\rm ej}$ (always >$P_{\rm ig}$), to eject the CE, and the orbital period range for producing sdB+NS binaries is from $P_{\rm ej}$ to $P_{\textrm{tip}}$.

Figure 1 shows the evolutionary tracks of the donors. 
Filled dots and triangles indicate the positions of the onset of RLOF for the models that can produce sdB+NS binaries from stable RLOF and from CE ejection, respectively. The first point on each track is determined by $P_{\rm min}^{\rm i}$. 
All the models starting RLOF on MS or during HG produce sdB+NS binaries through stable RLOF. 
For those starting RLOF on the GB, sdB+NS binaries are also produced through stable RLOF when $M_{\rm d}^{\rm i}\le 2.5 \,M_\odot$, and the last position is determined by $P_{\rm max}^{\rm i}$. For models with $M_{\rm d}^{\rm i}\ge 3.2 \,M_\odot$, the sdB+NS binaries are produced through the CE ejection channel, and the first triangle on the GB is determined by $P_{\rm ej}$ or $P_{\rm ig}$.

  \begin{figure}
   \centering
   \includegraphics[width=\hsize]{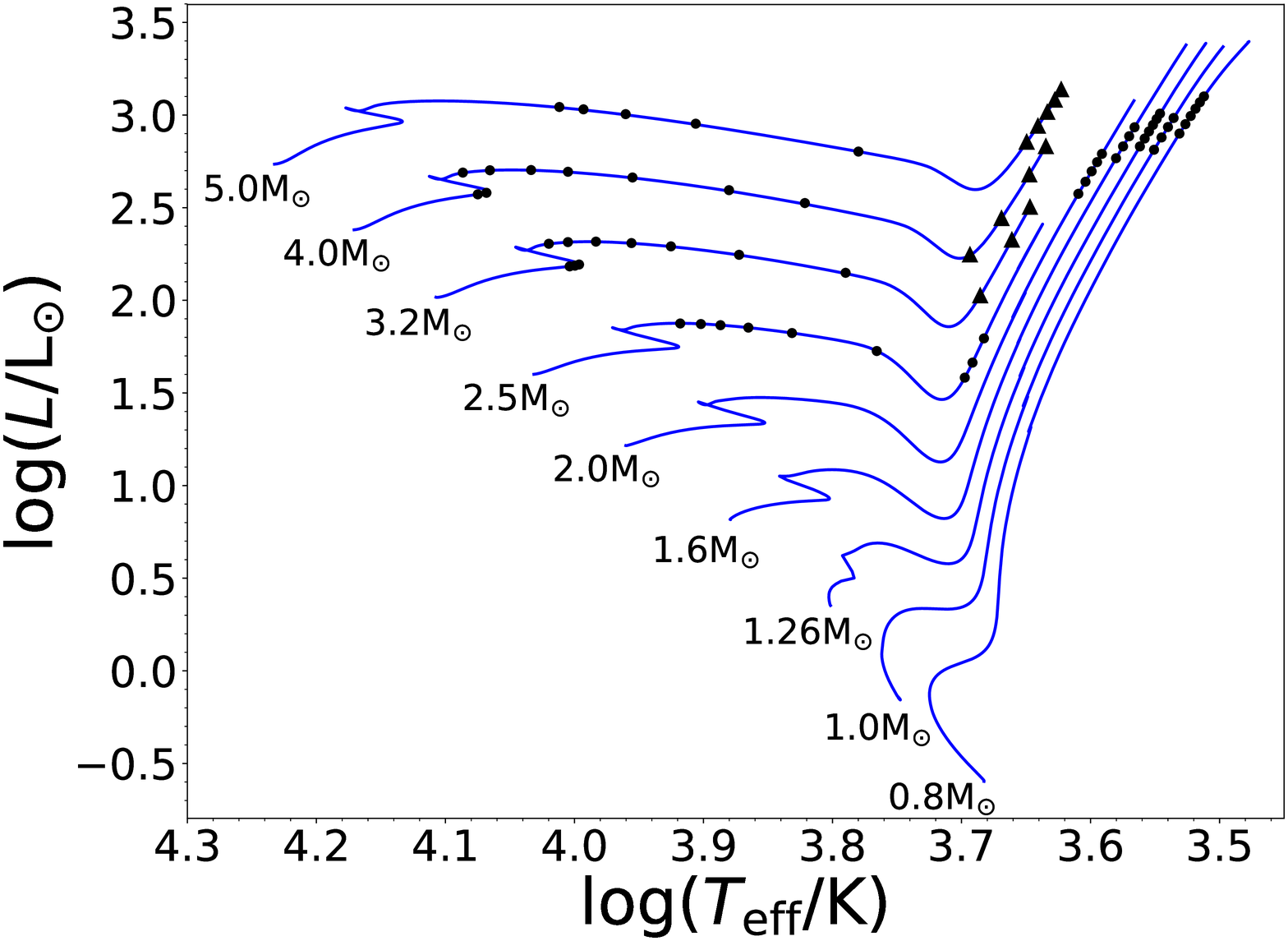}
      \caption{Evolutionary tracks of the donors on HRD. The positions at the onset of mass transfer are indicated for the models that can produce sdB+NS binaries through stable RLOF (filled dots) and through CE ejection (triangles). See text for details.}
         \label{hrd}
   \end{figure}
   
 \begin{figure*}
   \centering
   \includegraphics[width=\hsize,height=21cm]{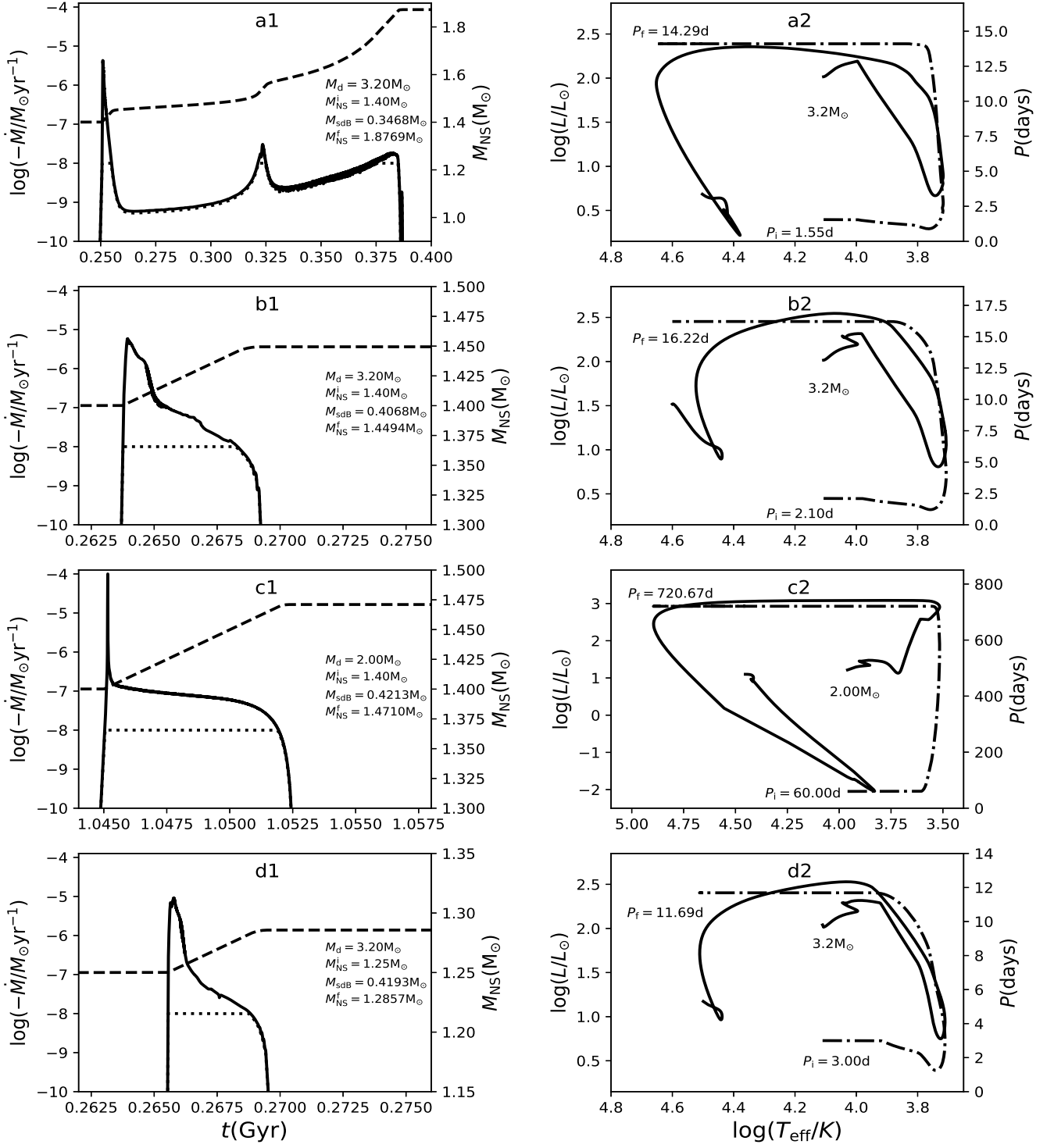}
      \caption{Details of the evolution for four representative cases in our study. Panels a to c show the models in which the donor starts mass transfer at different evolutionary phases, i.e., on the MS, during the HG, and on the GB, respectively, with an NS mass of $1.4M_\odot$, and panel d shows an alternative to the NS mass, i.e., $M_{\rm NS}^{\rm i}=1.25M_\odot$. For each of the models, the mass-transfer rate (solid line), the mass-accretion rate of the NS (dotted line), and the mass of the NS (dashed line) are shown in the left panel, and the corresponding evolutionary track (solid line) and orbital period change (dot-dashed line) are presented in the right panel. The initial and final component masses are indicated in the left panel, and the initial and final orbital periods , $P_{\rm i}$ and $P_{\rm f}$, are indicated in the right panel.}
         \label{examples}
   \end{figure*}

 \begin{figure*}
   \centering
  \includegraphics[width=\hsize]{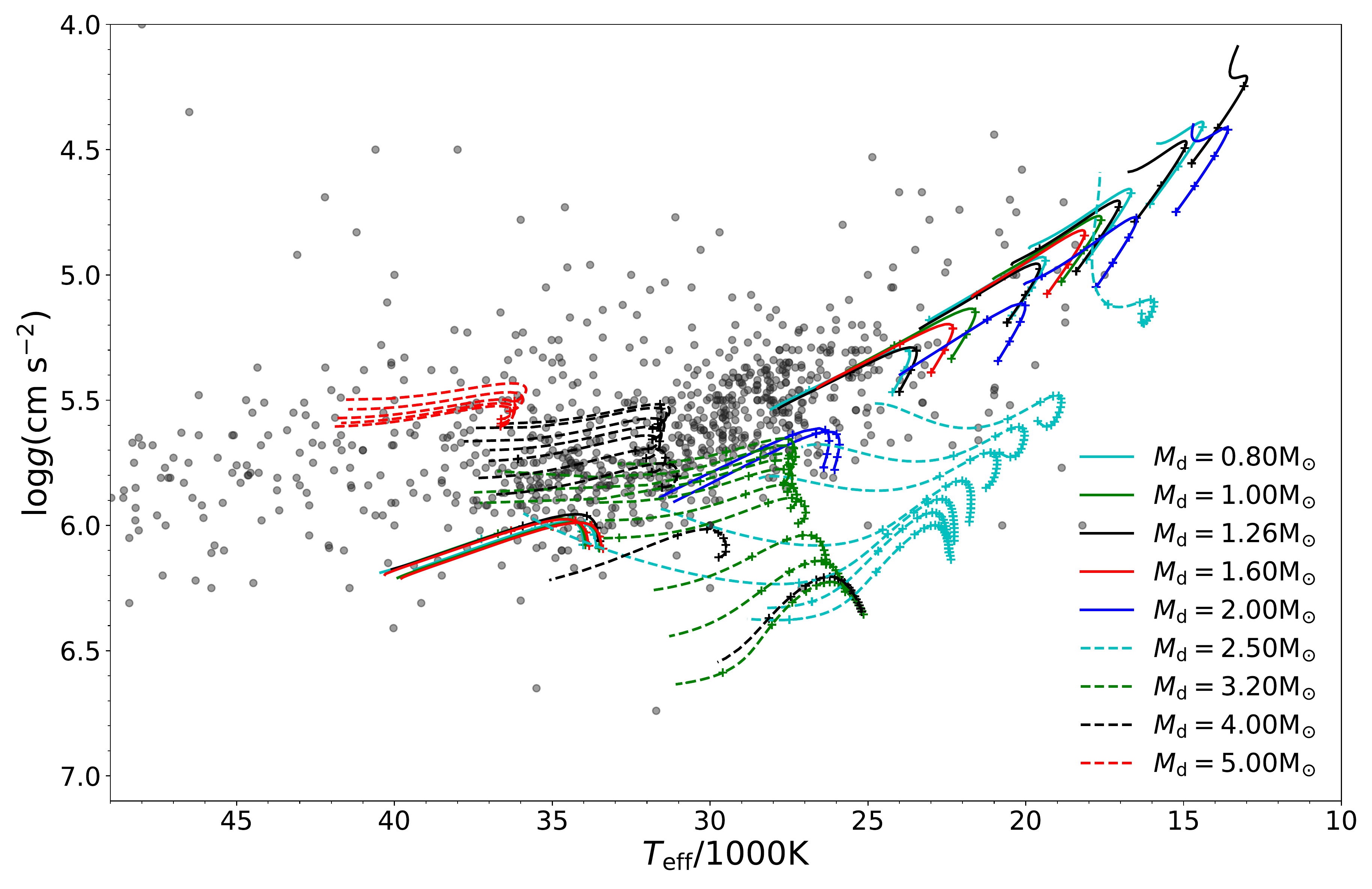}
   \caption {Effective temperature-gravity diagram for sdBs produced in our study. For clarity, only the He-core burning phase is presented here. The solid lines show those from degenerate He cores before its ignition ($M_{\rm d}^{\rm i} \le 2.0 \,M_\odot$), and the dashed lines show those from non-degenerate He cores  ($M_{\rm d}^{\rm i} \ge 2.5 \,M_\odot$).  Different colors represent different donor masses as indicated in the figure. The age differences between adjacent crosses are $2.5\times 10^{7}$ yr for each evolutionary track. The dots show known hot subdwarfs \citep{Geier2016a}.} 
              \label{tg}%
\end{figure*}

 \begin{figure}
   \centering
   \includegraphics[width=\hsize]{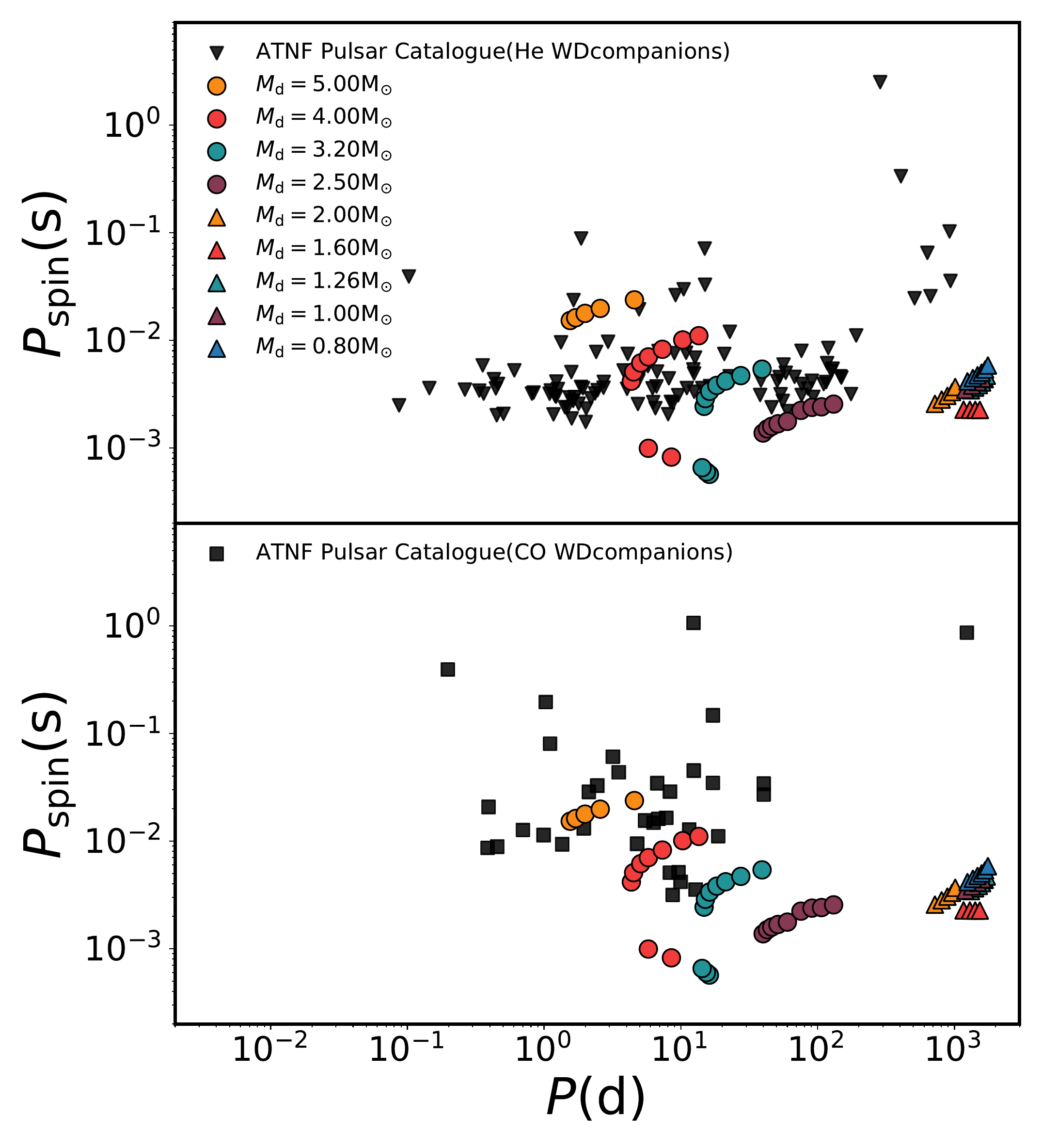}    
   \caption{Diagram of orbital period - spin period for sdB+NS binaries. Different colors  represent different initial donor masses.  Binary radio pulsars in the Galactic disk from the ATNF Pulsar Catalogue \citep{Manchester2005} are shown in the figure for comparison, where triangles (in the upper panel) show those with a He WD companion, and squares show those of CO companions. }
              \label{pp}%
\end{figure}

\section{Stable mass transfer channel}  

\subsection{Production of sdB+NS binaries}
Fig.1 shows that He can only be ignited after the mass transfer when the donors start mass transfer close to the tip of the RGB if $M_{\textrm{d}}^{\textrm{i}}\leq 2.0 \,M_\odot$, consistent with results of \citet{1999A&A...350..928T}. For more massive donors, the He core can be ignited after the mass transfer if the donor starts mass transfer earlier, that is, during early HG for $M_{\textrm{d}}^{\textrm{i}}= 2.5 \,M_\odot$ and 
near the end of the MS for $M_{\textrm{d}}^{\textrm{i}}= 3.2 \,M_\odot$ and $4.0M_{\odot}$.
For the case of $M_{\textrm{d}}^{\textrm{i}} = 5.0 \,M_\odot$, if the initial orbital period is short and mass transfer begins earlier than shown by the first dot in the figure, the total angular momentum $J$ becomes negative as a result of the mass loss from the system, leading to a dramatic shrinkage of the orbit and the merger of the binary.
 
In Fig. 2 we present the evolution in detail for four representative examples in our calculations. The first three are for the models in which the donor starts mass transfer at various evolutionary phases, that is, on the MS, during the HG and on the GB, respectively, with an NS mass of $1.4 \,M_\odot$. The fourth example is an alternative to the NS mass, that is, $M_{\textrm{NS}}^{\textrm{i}} = 1.25 \,M_\odot$. The mass transfer rate, the mass accretion rate of the NS, and the mass of the NS are shown in the left panels, and the corresponding evolutionary tracks and variations in orbital periods are presented in the right panels. The initial and final parameters for each model are indicated in each panel.

There are two peaks for mass transfer rate when the donor starts mass transfer near the end of the MS, as shown in panel (a1). The first peak appears when the donor is still on the MS, while the second peak emerges as the donor evolves along the GB. Most of the material of the donor is transferred onto the NS surface with a sub-Eddington rate, and a large portion of the material has then been accreted by the NS, leading to a significant increase in NS mass. For the other cases, as shown in panels (b1) to (d1), however, the mass transfer rate is usually higher than the Eddington rate. As a result, most of the transferred mass cannot be accreted by the NS and is lost from the system. The NS accordingly increases little in mass.

Table 1 summarizes the properties of the sdB+NS systems from our simulations, such as the initial mass of the donor ($M_{\rm d}^{\rm i}$), the orbital period ($P_{\rm f}$), component masses ($M_{\rm sdB}$, $M_{\rm NS}$), the total hydrogen mass ($M_{\rm H}$), the RV semi-amplitude ($K$), and the spin period of the NSs $(P_{\rm NS}$, see Sect. 3.4). The value of $K$ is calculated from binary mass function \citep{Geier2011}:

\begin{equation}
      f_{\mathrm{m}}=\frac{M_{\mathrm{NS}}^{3}{\sin^{3} i}}{(M_{\mathrm{NS}}+M_{\mathrm{sdB}})^{2}}=\frac{PK^{3}}{2\pi G}\,,
   \end{equation}
where the masses are in units of solar mass, $P$ is in unit of days, and $G$ is the gravitational constant in unit of cgs.  $K$ has a maximum theoretical value by assuming an inclination angle ($i$) of $90^{\circ}$. 

Table 1 shows that the earlier the mass transfer starts, the lower the produced sdB mass and the higher the NS mass. This is easily understood from binary evolution. In general, a He core mass (the remnant mass after mass transfer) increases with the postponement of RLOF-starting time for similar adoptions for binary evolution. If the mass transfer starts early, the mass transfer process is relatively moderate and lasts for a long time, as shown in Figs. 2a to 2c. The NS then has an opportunity to accrete more material, in comparison to the case where the donor starts mass transfer in later evolutionary phases, and increases its mass significantly.

Table 1 also shows that if the mass transfer starts when the donors are during HG or later, the final orbital period increases and the maximum value of $K$ decreases monotonically with the mass transfer starting time. However, if an sdB+NS system can be produced when the donor starts mass transfer near the end of the MS, that is,. $M_{\textrm{d}}^{\textrm{i}} = 3.2 \,M_\odot$ and $4.0 \,M_\odot$, the shortest final orbital period is obtained at the MS turn-off point rather than the first point (determined by $P_{\textrm{min}}^{\textrm{i}}$) on the evolutionary tracks. For these models, the differences in initial orbital periods are small, but the mass loss (then angular momentum loss) increases significantly with the postponement of RLOFstarting time and offsets such differences, resulting in the shortest orbital period when the donor starts RLOF at the turn-off of the MS. Accordingly, the highest value of $K$ also appears at or near this point. We discuss the value of $K$ in more detail in Sect.4, together with the value from the CE ejection channel.

For a donor star with a mass lower than $2.0 \,M_\odot$, the final orbital period can reach up to $\sim$1800d because we adopted the Ritter scheme for the mass transfer mode, similar to the atmospheric RLOF discussed in \citet{Chen2013}. Our results are consistent with those estimated by \citet{Chen2013} ($\sim$1600d).
\subsection{Evolution of sdBs}

We have traced the evolution of the produced sdBs and compared them with the known sdBs \citep{Geier2016a} on the effective temperature-gravity (in logarithmic) diagram, as shown in Fig.3. For clarity, we only show the He core-burning phase in the figure. Crosses on each of the evolutionary tracks indicate the  evolution timescale of sdBs. The age differences between adjacent crosses are $2.5\times 10^{7}$ yr.  Although the samples from the observations are not sdB+NS systems, the structures of sdBs produced from a given donor mass are similar regardless of the companion types \citep{Han2002}. This means thatfF the information indicated in Fig.3 holds in general for the sdB population from the stable RLOF channel, including the following:

(i) The sdBs produced from $M_{\textrm{d}}^{\textrm{i}}\leq 2.0 \,M_\odot$ have masses around $0.47 \,M_\odot$ and are separated into two groups. One group gathers on the highest temperature end of the EHB. These sdBs have experienced a delayed He flash before central He burning, and nearly no H-rich envelope has left after the flashes. The other group has a significant H-rich envelope and scatters around the EHB, leaving a gap between the two groups. This is very similar to the case of the CE ejection channel \citep{Xiong2017}.

(ii) A mass of sdBs that is lower or higher than the canonical mass ($0.47 \,M_\odot$) can be produced by more massive donors, consistent with the results of \citet{Han2002}. A remarkable phenomenon is that some of these products are the gap left by the products from less massive progenitors. It becomes more interesting, moreover, because the place of the gap coincides with the location of blue hook stars that were discovered in some globular clusters \citep{Lei2015}, where stars with relatively high masses (the exact mass is determined by the age of the cluster) are believed to have completed their evolution and no longer contribute to sdB formation. A detailed study of this problem is beyond the scope of this paper.

(iii) It is quite clear that sdBs with low masses have much longer lifetimes than those with high masses. However, only a few objects have been found in the location occupied by the low-mass sdBs  ($M_{\textrm{sdB}} < 0.47 \,M_\odot$). The discrepancy could be caused by selection effects, for instance, the luminosity of such sdBs is significantly lower than that of massive sdBs and could be below the detection limit for MUCHFUSS. The disadvantages of the initial mass function (IMF), that is, the number of stars with exponentially increasing mass  
may also lead to this discrepancy since they have fewer progenitors. If the discrepancy were to arise from the distribution of IMF, there should also be fewer high-mass sdBs, although many observational samples are obtained in locations that the evolutionary tracks of high-mass sdBs pass through. As Fig.3 shows, many of the observed sdBs lie outside the He-core burning region given by our models. These objects likely have CO cores.

\subsection{Effects of NS mass and accretion efficiency}

NS accretion is a complex process, and many microphysics in this process need to be understood better. 	
We here assumed a coefficient $\varepsilon$ to describe the NS accretion and set it to 0.9 for sub-Eddington accretion for the simulation above. However, the NS accretion efficiency has been demonstrated to be much lower than 0.5-0.9 in many binary pulsar systems. For example, \citet{2012MNRAS.423.3316A} reproduced the evolutionary history of the pulsar-WD binary PSR J1738+0333 and found that more than 60\% of transferred matter from the WD progenitor has been lost from the system, and the value of $\varepsilon$ may not exceed 0.5. To examine the effect of this parameter, we adopted two other values here, 0.7 and 0.5, to show the evolutionary consequences for binaries with 
a $3.2 \,M_\odot$ donor and a $1.4 \,M_\odot$ NS. The results are shown in Table 3.

The $M_{\mathrm{NS}}$ decreases significantly (about $0.2 \,M_\odot$) with decreasing $\varepsilon$ when the donor starts RLOF on the MS, but changes little when the RLOF starts in later evolutionary phases. 
The reason is  that $\varepsilon$ only describes the mass fraction accreted by an NS when the mass transfer rate is sub-Eddington, which occurs and takes a long time when the donor starts RLOF on the MS (see Fig.2). 
With increasing of $P_{\mathrm{i}}$, the RLOF starts later and the mass transfer rate increases dramatically and rapidly exceeds the Eddington rate. In this case, the NS accretes material with the Eddington rate and is not affected by the parameter $\varepsilon$. The orbital period, the RV semi-amplitude, and the spin of the NS vary with $M_{\mathrm{NS}}$ accordingly, but the sdB mass and the parameter space for producing the sdB+NS binaries change little. 

As described in Sect.1, an NS born from the ECSNe or AIC has a mass of about $1.25 \,M_\odot$  and a weak natal kick.  A binary consisting of such an NS is then more likely to survive SN explosions. Hence, we calculated the evolutions for binaries consisting of a $1.25 \,M_\odot$ NS and a $3.2 \,M_\odot$ MS star with $\varepsilon=0.9$. The results are listed in Table 4. An example of such systems is shown in panel (d) of Fig.2, where RLOF starts when the donor is in the HG. The mass transfer and accretion processes are similar to those shown in panel (b), which is for the same donor star starting RLOF during the HG, but with a $1.4M_\odot$ NS. In comparison to the case of $1.4M_\odot$ NS, the initially higher mass ratio ($M_{\rm d}/M_{\rm NS}$) causes a relatively higher mass transfer rate, then a shorter timescale, and more material has been lost from the system, resulting in shorter orbital periods, as presented in Table 4. Although the final NS mass is lower than that from $1.4M_\odot$ NS, the RV semi-amplitude $K$ increases slightly due to their shorter orbital periods. The parameter space for producing sdB+NS binaries is hardly affected by the NS mass.

\subsection{NS spin}

Since the accreted material carries angular momentum, the NS is spun up during the accretion process. 
This recycled scenario has been successfully used to explain the formation of pulsars and millisecond pulsars \citep{Alpar1982,Radhakrishnan1982,Bhattacharya1991}. The spin period of an NS after accretion can be roughly estimated according to the formula of \citet{Tauris2012}:
\begin{equation}
      {\Delta M_{\mathrm{eq}}=0.22M_{\odot }\frac{(M_{\rm NS}/M_{\odot })^{1/3}}{P_{\mathrm{ms}}^{4/3}}}\,,
   \end{equation}
where $\Delta M_{\mathrm{eq}}$ is the accreted mass and $P_{\mathrm{ms}}$ is the equilibrium spin period in units of milliseconds.

The spin periods of NSs in the produced sdB+NS binaries are presented in the last column of Tables 1 to 4. 
Most of the NSs are spun-up to be millisecond pulsars. 
For instance, a system with $M_{\mathrm{d}}=3.2M_{\odot }$, $M_{\mathrm{NS}}^{\rm i}=1.4M_{\odot }$, and $P^{\mathrm{i}}=1.60\mathrm{d}$ has an equilibrium spin period of 1.83 ms after the accretion process. This means that the pulsar signals could be a characteristic for sdB+NS binaries produced from the stable RLOF channel.

For the stable RLOF, the donor evolves directly to a He WD
 if He cannot be ignitied in the core after the mass transfer. 
 Meanwhile, the sdB+NS systems will evolve to CO WDs after central He has been exhausted.  
 We therefore compared our results and the observed binary pulsars with WD companions from the ATNF Pulsar Catalogue \citep{Manchester2005}. The results are presented in Fig. 4. 
 
Fig.4 shows that the models from $M_{\rm d}^{\rm i} \ge 2.5M_\odot$ are generally consistent with observations of binary pulsars. However, there are significant differences for systems with relatively long orbital periods ($P>>100$ d). The observed spin period is longer than that in our study by more than one order of magnitude. 
The products of such long-orbit-period binaries are from stable RLOF  when the donors are on the GB, where the mass transfer rate most of the time reaches the super-Eddington rate and the NS increases in mass with the Eddington accretion rate, as shown in panel (c) of Fig.3. The discrepancy between observations and our results indicates that the real accretion efficiency could be far lower than the Eddington rate in this case. 
Nevertheless, the NSs in sdB binaries can be spun up to be millisecond pulsars if the progenitors for sdBs are relatively massive (having non-degenerate cores) and start mass transfer on the MS or during the HG. 

 \begin{figure}
   \centering
   \includegraphics[width=\hsize]{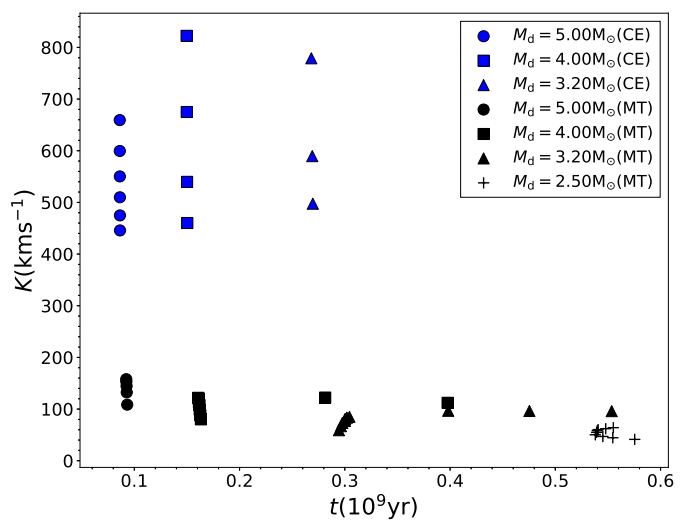}
   \caption{Formation age of sdB+NS binaries vs. RV semi-amplitude. The blue symbols show binaries from the CE ejection channel, and the black symbols represent binaries from stable RLOF. Different shapes show different initial donor masses as indicated in the plot.}
              \label{K-t}%
    \end{figure}
    
 \begin{figure}
   \centering
   \includegraphics[width=\hsize]{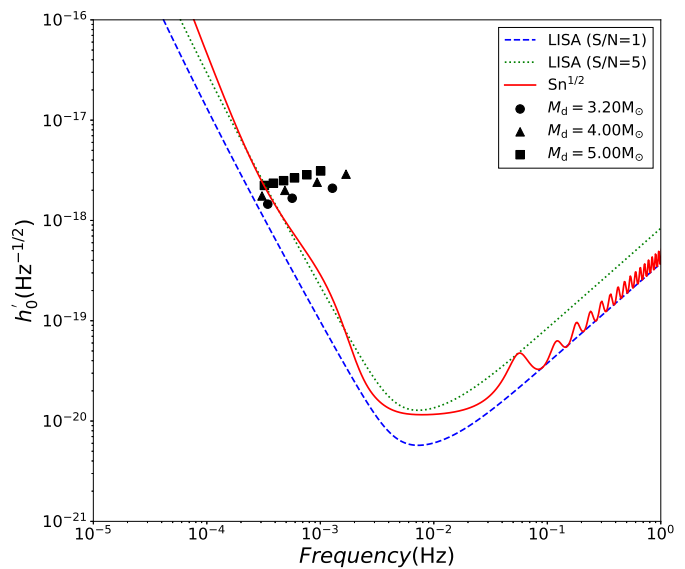}
   \caption{Root spectral density of GW strain as a function of GW frequency. Black symbols represent the GW signal of sdB+NS systems from the CE ejection channel, where $\tilde{D}$=10kpc, $T=4yr$ are assumed. The dashed and dotted lines plot the sensitivity curves of LISA with an S/N=1 and S/N=5, respectively \citep{2016PhRvD..93b4003K}, and the solid line shows the full sensitivity curve combining the instrument noise and Galactic confusion noise \citep{2018arXiv180301944C}.}
              \label{K-t}%
    \end{figure}
\section{Common envelope ejection channel}
The CE ejection channel is generally considered as a major scenario for the formation of sdB binaries. It is an unsolved puzzle whether the NS accretes material during the spiraling-in process in the CE evolution.
The mass increase should be very small if accretion indeed occurs because the timescale for the CE process ($\sim$1000 yr)\footnote{It was suggested that hyper-Eddington accretion onto an NS is possible if the gravitational energy released in accretion is lost by neutrino \citep{Colgate1971,Zel'dovich1972}, and this may be the case for the NS in CE \citep{Chevalier1993}. The accretion rate in this case may be as high as $\sim 0.1M_{\odot}/yr$, which may result in the collapsing of the NS into a BH inside the CE. Although some arguments have been made in aid of this hypothesis, the possibility of hyper-Eddington accretion has been included in several population synthesis studies to account for evidence of a diminishing population of NS+NS binaries in favor of NSs in pairs with low-mass BHs \citep{Portegies1998,Belczynski2002}.} is very short. Here we assumed that there is no accretion for NSs in the CE process and obtained the final orbital periods from Eq.(1) by setting $\alpha_{\rm th}=\alpha_{\rm CE}=1$. This gives the longest orbital periods for sdB binaries formed in this way, and the value of $K$ was obtained accordingly. The results are summarized in Table 5. Because the short orbital periods after the CE process are short, $\text{approximately}$ some hours, 
the value of $K$ is dramatically higher than that from the stable RLOF channel.

However, the progenitors for sdBs in this way are relatively massive and should have evolved off the MS if they are born in the Galactic halo. Fig. 5 presents the formation age of sdBs versus RV semi-amplitude. There might be small difference for the ages if we were to consider the evolution before the NS formation, but it does not affect the discussion below. The systems from $M_{\mathrm{d}}^{\mathrm{i}}\leq 2.0\mathrm{M_{\odot }}$ are not included in Fig. 5 since they have very low $K$ and are much older than 0.6 Gyr (the maximum value for the x-axis). 
Obviously, the sdB formation age decreases with increasing progenitor masses. 
It should be younger than about 0.3 Gyr for systems with $K$ up to $\sim 500 \rm km\ s^{-1}$ or more. 
This is probably the reason for that the MUCHFUSS project has not discovered a an sdB+NS system so far.
 Their candidates are from the SDSS, which covers high Galactic latitudes, and most of the candidates likely come from the halo and are old.  
 
The very short orbital periods for sdB+NS systems produced in this way (on the order of hours, as shown in Table 5) and the relatively high component masses make such systems potential strong gravitational wave (GW) sources in the frequency range of $10^{-4}$ to 1 Hz and could be resolved by the future Laser Interferometer Space Antenna (LISA). The root
spectral density of GW strain can be estimated by \citep{baz}


    \begin{equation}
\begin{split}
{h}'_{0}=h_{0}\sqrt{T}=\frac{2G^{2}M_{1}M_{2}}{\tilde{D}a}\sqrt{T}\approx & 8\times 10^{-13}/\mathrm{H_{z}}^{1/2}(\frac{M_{1}}{M_{\odot }})(\frac{M_{2}}{M_{\odot }})\\
&\times (\frac{\mathrm{kpc}}{\tilde{D}})(\frac{\mathrm{km}}{a})(\frac{T}{\mathrm{yrs}})^{1/2},
\end{split}
    \end{equation}
where $G$ is the gravitational constant and $a$ is the orbital separation, $M_{1}$ and $M_{2}$ are the mass of sdB and NS, respectively, $\tilde{D}$ is the distance to the binaries, and $T$ is the integration time for observations. The results are presented in Table 4 by assuming $\tilde{D}$=10kpc and $T=4yr$, which corresponds to the mission lifetime of LISA \citep{2016PhRvD..93b4003K}. The GW frequency and corresponding root spectral density for each system, as well as the LISA sensitivity curves, are presented in Fig. 6. For the LISA sensitivity curves, we chose $5\times 10^{6}\mathrm{km}$ as the arm length $L$ with different signal-to-noise ratio (S/N) values \citep{2016PhRvD..93b4003K}. \citet{2018arXiv180301944C} showed a sensitivity curve of LISA by combining the noises of the instrument and the Galactic confusion based on an analytic model. The curve is also included in Fig. 6 for comparison.  
Almost all the sdB+NS systems produced in this way are far beyond the sensitivity curves of LISA and could be resolved by LISA in future.

Adopting the IMF of \citet{1979ApJS...41..513M} and a constant star formation rate of $5M_{\odot}yr^{-1}$ for the Galaxy,
we obtain a birthrate of sdB+NS systems from the CE ejection channel (potential GW sources) of $\sim2\times 10^{-5}yr^{-1}$, 
and the number is about 20, if the natal kicks of NSs follow the Maxwellian distribution \citep{1997MNRAS.291..569H} with a velocity scatter $\sigma=190 {\rm km\ s^{-1}}$. 
The natal kicks would change the results significantly, and a detailed study will be given in the next paper.

However, the lifetimes of such systems is short, on the orders of $\sim$ Myr, which means that the sdB stars will overflow their Roche lobes soon after the CE ejection, very close to zero age (the lifetime for sdBs is $\sim 10^{8}$yr). The following mass transfer process is probably stable and lasts for a relatively long lifetime. The systems in this evolutionary stage may be recognized as ultracompact X-ray binaries with sdB donors. The sdB donors may extinguish the He-burning due to mass loss during the mass transfer process and evolve to be He-WDs.

\section{Discussion and conclusions}

We investigated the formation of sdB+NS systems from a series of MS+NS binaries. 
The parameter space for producing sdB+NS binaries was obtained for this study 
for the stable RLOF channel and the CE ejection channel. 
If the MS mass is initially lower than $2M_\odot$ , the mass transfer process from the donors to NSs is dynamically stable 
and the sdBs can only be formed when the donor starts mass transfer close to the tip of FGB. 
For more massive MS stars, however, the sdBs can be produced when the donor starts mass transfer during the HG or even earlier (near the end of the MS). 
If the donor is more massive than $3.2M_\odot$ and starts mass transfer on the GB, 
the mass transfer is assumed to be dynamically unstable, 
then a sdB+NS binary is formed if the CE can be ejected based on the standard energy budget, and the He core is ignited after the CE ejection.

The sdB+NS binaries from the stable RLOF generally have long orbital periods, as expected, 
ranging from several days to more than 1000 days, and the range of the orbital period moves toward the short-period side with increasing initial MS mass. The largest RV semi-amplitude is derived at about $150 {\rm km\ s^{-1}}$ for an MS star with initially $5M_\odot$ . We studied Population I stars here. With decreasing metallicity, the orbital periods from the stable RLOF are expected to be significantly shorter than the period shown in this paper, according to the study of \citet{Chen2013}. As a consequence, the value of $K$ increases and the resulting sdB+NS binaries at low metallicity could be discovered more easily. The sdB+NS systems that resulted from the CE ejection have very short orbital periods and then a large RV semi-amplitude (the value of $K$ may be up to $800 \rm km\ s^{-1}$). Such systems are born in very young populations, that is, younger than $\sim$ 0.3 Gyr, and are potential strong GW sources with frequencies between $10^{-4}-1$, and could be resolved by LISA in future.
Their lifetimes are very short ($\sim$ Myr) because of the GW radiation, which means that such systems must be rare and also hard to detected.

In the sdB+NS systems from the stable RLOF, almost all the NSs have been spun up to be millisecond pulsars in our study.
The observations of binary pulsars with WD companions indicates that 
the models from relatively massive MS stars ($M_{\rm d}^{\rm i} \ge 2.5M_\odot$) are generally consistent with the observations, but a discrepancy appears if the donors have lower initial masses. 
The sdB+NS binaries from the latter have long orbital periods ($P>>100$ d) and very short spin periods, 
more than one order of magnitude shorter than that of observed binary pulsars with long orbital periods.
This indicates that the real accretion efficiency could be far lower than we assumed when the mass transfer rate is dramatically higher than the Eddington rate, for instance, when mass transfer occurs when the donor is on the GB.
Nevertheless, the pulsar signal is likely a feature of sdB+NS systems that result from the stable RLOF channel. 

The NS accretion efficiency has significant effect on the final NS mass if the donor starts mass transfer on the MS, 
but little effect on the sdB mass and the parameter space for producing sdB+NS systems. In comparison to the case of $1.4M_\odot$ NS,  a $1.25M_\odot$ NS  leads to a significantly shorter final orbital period becausee the mass transfer rates induced by the higher initial mass ratio are relatively higher. Therefore more mass and then more angular momentum have been lost from the system, leading to a more dramatic shrinking of the orbit.
The X-ray irradiation from the NS to the donor during accretion may affect the detailed evolutionary process. This would be important and needs further studies. The numbers and statistical properties of binary parameters for sdB+NS binaries will be given from binary population synthesis in the next paper. 

\section{Acknowledgements} 

This work is partly supported by the Natural Science Foundation of China (Nos.11733008, 11422324, 11521303 ) and by Yunnan province (Nos. 2017HC018, 2013HA005). 

\begin{table*}
\caption{\label{t7}Evolutionary consequences in our simulation with different initial parameters.}
\centering
\begin{tabular}{cccccccc}
 \hline \hline
$M_\mathrm{d}^{\mathrm{i}}(\mathrm{M}_\odot)^{a}$&$P_\mathrm{i}(\mathrm{d})^b$ & $P_\mathrm{f}(\mathrm{d})^c$ & $M_\mathrm{H}(\mathrm{M}_\odot)^d$ & $M_\mathrm{NS}(\mathrm{M}_\odot)^e$ & $M_\mathrm{sdB}(\mathrm{M}_\odot)^f$ & $K(\mathrm{km\ s^{-1}})^g$ & $P_{\mathrm{NS}}(\mathrm{ms})^h$\\
\hline
&350 & 1246.95 & $6.8713\times 10^{-6}$ & 1.4371 & 0.4536 & 18.59 & 4.16 \\
&400 & 1365.15 & $8.6433\times 10^{-6}$ & 1.4338 & 0.4620 & 17.97 & 4.46 \\
0.8&450& 1477.21 & $2.3714\times 10^{-3}$ & 1.4311 & 0.4696 & 17.44 & 4.74 \\
&500 & 1583.21 & $0.7298\times 10^{-3}$ & 1.4289 & 0.4768 & 16.98 & 5.02 \\
&550 & 1678.92 & $1.2425\times 10^{-2}$ & 1.4264 & 0.4844 & 16.60 & 5.36 \\
&600 & 1759.72 & $1.8188\times 10^{-2}$ & 1.4235 & 0.4928 & 16.27 & 5.85 \\
\hline

& 220 & 1201.64 & $2.7696\times 10^{-5}$ & 1.4474 & 0.4507 & 18.91 & 3.47 \\
1.0 &     260 & 1346.65 & $7.6504\times 10^{-5}$ & 1.4419 & 0.4611 & 18.11 & 3.80 \\
&     300 & 1482.38 & $0.4308\times 10^{-2}$ & 1.4378 & 0.4704 & 17.45 & 4.10 \\
&     340 & 1606.51 & $1.0476\times 10^{-2}$ & 1.4342 & 0.4795 & 16.92 & 4.42 \\ \hline
 
&180 & 1320.79 & $0.0132\times 10^{-2}$ & 1.4480 & 0.4599 & 18.27 & 3.44 \\
&    200 & 1418.17 & $0.2459\times 10^{-2}$ & 1.4444 & 0.4667 & 17.78 & 3.64 \\
&   220 & 1510.63  & $0.6860\times 10^{-2}$ & 1.4413 & 0.4731 & 17.35 & 3.84 \\
1.26   &   240 & 1595.36 & $1.1396\times 10^{-2}$ & 1.4382 & 0.4797 & 16.98 & 4.07 \\
&   260 & 1668.25 & $1.6462\times 10^{-2}$ & 1.4348 & 0.4871 & 16.66 & 4.37 \\
&   280 & 1727.33 & $2.2124\times 10^{-2}$ & 1.4316 & 0.4955 & 16.41 & 4.69 \\ \hline

& 120 & 1167.79 & $2.6212\times 10^{-5}$ & 1.4542 & 0.4496 & 19.14 & 2.27 \\
1.6 &    140 & 1298.25 & $5.2976\times 10^{-5}$ & 1.4480 & 0.4590 & 18.38 & 2.26 \\
&   160 & 1421.06 & $3.3633\times 10^{-3}$ & 1.4434 & 0.4676 & 17.75 & 2.26 \\
&   180 & 1535.28 & $9.0438\times 10^{-3}$ & 1.4394 & 0.4757 & 17.23 & 2.26 \\ \hline

& 60 & 720.67 & $0.4354\times 10^{-3}$ & 1.4710  & 0.4213 & 22.84 &2.57\\
& 70 & 807.16 & $0.4539\times 10^{-3}$ & 1.4628 & 0.4291 & 21.87 &2.82\\
2.0& 80 & 889.03 & $4.4677\times 10^{-3}$ & 1.4564 & 0.4362 & 21.08 &3.05\\
& 90& 962.22 & $1.0397\times 10^{-2}$ & 1.4505 & 0.4437 & 20.44 &3.31\\
& 100 & 1015.02 & $1.7280\times 10^{-2}$  &1.4437 & 0.4539 & 19.96 & 3.69\\ \hline

& 1.8 & 39.92 & $0.9662\times 10^{-3}$ & 1.5669 & 0.3152 & 64.04&1.37 \\
& 2.0 & 42.93 & $1.1730\times 10^{-3}$ &  1.5478 & 0.3202 & 62.06 &1.50\\
& 2.2 & 46.17 & $1.2561\times 10^{-3}$ &  1.5369 & 0.3236 & 60.30 &1.59 \\
& 2.5 & 51.27 & $1.4726\times 10^{-3}$ &  1.5270  & 0.3270  & 57.99 &1.59\\
2.5& 3.0  & 60.09 & $1.9198\times 10^{-3}$ &   1.5184 & 0.3309 & 54.79&1.77 \\
& 4.0   & 75.49 & $2.9492\times 10^{-3}$ &  1.4860  & 0.3363 & 50.18&2.23\\
& 5.0 & 90.94 &$3.6463\times 10^{-3}$ &         1.4787 & 0.3403   &     46.99& 2.28\\
& 6.0& 106.75& $4.6129\times 10^{-3}$ & 1.4775& 0.3433& 44.48& 2.41\\
& 8.0&130.87& $9.8698\times 10^{-3}$ &1.4716&   0.3550&         41.30&2.56\\ \hline

& 1.45 & 16.15 & $0.3699\times 10^{-3}$ &  1.9872 & 0.3165 & 95.97 &0.57\\
& 1.50  & 15.36 & $0.6398\times 10^{-3}$ & 1.9432 & 0.3294 & 96.30 &0.60 \\
& 1.55 & 14.29 & $1.0820\times 10^{-3}$ & 1.8769 & 0.3468 & 96.67 &0.66\\
& 1.60  & 14.78 & $1.9653\times 10^{-3}$ & 1.4765 & 0.3770  & 84.90 &2.43 \\
& 1.80  & 15.10  & $2.2697\times 10^{-3}$ & 1.4610  & 0.3928 & 83.40 &2.88 \\
3.2& 2.10  & 16.22 & $2.6429\times 10^{-3}$ & 1.4494 & 0.4068 & 80.72 &3.36\\
& 2.50  & 18.30  & $2.9440\times 10^{-3}$ & 1.4417 & 0.4162 & 77.08 &3.81\\
& 3.00    & 21.26 & $3.1447\times 10^{-3}$ & 1.4368 & 0.4219 & 73.05 &4.19\\
& 4.00    & 27.43 & $3.4374\times 10^{-3}$ & 1.4315 & 0.4277 & 66.84 &4.70\\ 
& 6.00  &39.22& $3.7649\times 10^{-3}$ &1.4261  &0.4367 &       59.04&5.40\\
\hline
& 1.5     & 8.52    & $0.4569\times 10^{-3}$    & 1.7437     & 0.3198   & 112.16 &0.82  \\
& 1.6     & 5.78    & $0.7580\times 10^{-3}$    & 1.6633     & 0.3867   & 122.29&0.99   \\
& 1.7     & 4.34    & $2.5695\times 10^{-3}$    & 1.4370      & 0.4733   & 121.84 &4.17  \\
& 2.0       & 4.51    & $3.0979\times 10^{-3}$    & 1.4283     & 0.4993   & 118.84 &5.09  \\
4.0& 2.5     & 5.08    & $3.6861\times 10^{-3}$    & 1.4219     & 0.5222   & 113.07 &6.16  \\
& 3.0       & 5.78    & $4.0875\times 10^{-3}$    & 1.4184     & 0.5340    & 107.73 &7.02  \\
& 4.0       & 7.32    & $4.6059\times 10^{-3}$    & 1.4148     & 0.5462   & 99.030 &8.26  \\
&6.0&   10.31& $5.2752\times 10^{-3}$ & 1.4113& 0.5612&         87.78&10.09\\
&8.0&   13.52&  $5.4764\times 10^{-3}$ &1.4100& 0.5654&80.04&11.05\\

\hline\hline
\end{tabular}
\tablefoot{(a) Initial mass of the MS star; (b) initial period of the systems; (c) final period of the systems; (d) total hydrogen mass of sdBs; (e) final mass of the NS; (f) mass of sdBs; (g) RV semi-amplitude of the systems; (h) spin period of the NS.}
\end{table*}

\begin{table*}
\caption*{\label{t7}Table 1: Continued}
\centering
\begin{tabular}{cccccccc}
 \hline \hline
$M_\mathrm{d}^{\mathrm{i}}(\mathrm{M}_\odot)^{a}$&$P_\mathrm{i}(\mathrm{d})^b$ & $P_\mathrm{f}(\mathrm{d})^c$ & $M_\mathrm{H}(\mathrm{M}_\odot)^d$ & $M_\mathrm{NS}(\mathrm{M}_\odot)^e$ & $M_\mathrm{sdB}(\mathrm{M}_\odot)^f$ & $K(\mathrm{km\ s^{-1}})^g$ & $P_{\mathrm{NS}}(\mathrm{ms})^h$\\
\hline
& 4.5                        & 1.55                       & $5.3983\times 10^{-3}$                         & 1.4065                       & 0.6914                        & 157.91  &15.28                     \\
& 5.0                          & 1.70                        & $5.6714\times 10^{-3}$                        & 1.4060                        & 0.6964                        & 152.85  &16.23                     \\
5.0& 6.0                          & 1.99                       & $6.0758\times 10^{-3}$                       & 1.4053                       & 0.7049                        & 144.60   &17.81                     \\
& 8.0                          & 2.57                       & $6.5451\times 10^{-3}$                        & 1.4046                       & 0.7150                         & 132.33   &19.80                    \\
& 15.0                         & 4.57                       & $7.4858\times 10^{-3}$                      & 1.4036                       & 0.7336                        & 108.55 &23.79                      \\
  \hline  \hline  
\end{tabular}
\tablefoot{(a) Initial mass of the MS star; (b) initial period of systems; (c) final period of systems; (d) total hydrogen mass; (e) final mass of the NS; (f) mass of sdBs; (g) RV semi-amplitude of systems; (h) spin period of the NS.}
\end{table*}

\begin{table*}
\caption{\label{t7} Evolutionary consequences of a $3.2 \,\mathrm{M_\odot}$ donor star with various NS accretion efficiencies.  }
\centering 
\begin{tabular}{ccccccccc}
 \hline \hline
${\varepsilon}$ &$M_\mathrm{d}^{\mathrm{i}}(\mathrm{M}_\odot)^{a}$& $P_\mathrm{i}(\mathrm{d})^b$ & $P_\mathrm{f}(\mathrm{d})^c$ & $M_\mathrm{H}(\mathrm{M}_\odot)^d$ & $M_\mathrm{NS}(\mathrm{M}_\odot)^e$ & $M_\mathrm{sdB}(\mathrm{M}_\odot)^f$ & $K(\mathrm{km\ s^{-1}})^g$ &$P_{\mathrm{NS}}(\mathrm{ms})^h$\\
\hline

&& 1.45&17.17 & $0.4102\times 10^{-3}$ & 1.8727 & 0.3173 & 91.66& 0.66\\
&&1.50& 16.12 & $0.6335\times 10^{-3}$ & 1.8484 & 0.3299 & 92.72&  0.68\\
&&1.55& 14.91 & $1.0380\times 10^{-3}$ & 1.8032 & 0.3468 & 93.65&  0.74\\
&&1.60& 14.81 & $1.9376\times 10^{-3}$ & 1.4745 & 0.3768 & 84.80&  2.48\\
&& 1.70 & 14.95 & $2.1019\times 10^{-3}$ & 1.4656 & 0.3853 & 84.03&  2.73\\
0.7 &3.2& 1.80 & 15.15 & $2.2478\times 10^{-3}$ & 1.4592 & 0.3926 & 83.27&  2.94\\
&& 1.90 & 15.44 & $2.4361\times 10^{-3}$ & 1.4544 & 0.3989 & 82.43&  3.13\\
&& 2.00 & 15.77 & $2.5582\times 10^{-3}$ & 1.4508 & 0.4004 & 81.71&  3.29\\
&& 2.10 & 16.21 & $2.6480\times 10^{-3}$ & 1.4479 & 0.4071 & 80.72&  3.44\\
&& 2.50 & 18.30 & $2.9712\times 10^{-3}$ & 1.4408 & 0.4160 & 77.06&  3.88\\
&& 3.00 & 21.22 & $3.2089\times 10^{-3}$ & 1.4359 & 0.4221 & 73.05&  4.26\\
&& 4.00 & 27.27 & $3.4972\times 10^{-3}$ & 1.4313 & 0.4289 & 66.95&  4.70\\    
 \hline
&& 1.45 & 18.27 & $0.4216\times 10^{-3}$ & 1.7517 & 0.3186 & 87.18 &0.81\\
&& 1.50 & 17.16 & $0.6488\times 10^{-3}$ & 1.7346 & 0.3309 & 88.29 &0.84\\
&& 1.55 & 15.69 & $1.0592\times 10^{-3}$ & 1.7110 & 0.3475 & 89.93 &0.88\\
&& 1.60 & 14.86 & $1.9308\times 10^{-3}$ & 1.4707 & 0.3768 & 84.60 &2.58\\
&& 1.70 & 14.99 & $2.1026\times 10^{-3}$ & 1.4621 & 0.3854 & 83.86 &2.84\\
0.5 &3.2& 1.80 & 15.19 & $2.2254\times 10^{-3}$ & 1.4560 & 0.3926 & 83.11 &3.07\\
&& 1.90 & 15.48 & $2.2539\times 10^{-3}$ & 1.4516 & 0.3984 & 82.30 &3.26\\
&& 2.00 & 15.81 & $2.4091\times 10^{-3}$ & 1.4480 & 0.4034 & 81.48 &3.44\\
&& 2.10 & 16.24 & $2.2544\times 10^{-3}$ & 1.4455 & 0.4071 & 80.58 &3.58\\
&& 2.50 & 18.34 & $2.6811\times 10^{-3}$ & 1.4387 & 0.4161 & 76.95 &4.03\\
&& 3.00 & 21.26 & $2.9630\times 10^{-3}$ & 1.4341 & 0.4222 & 72.98 &4.43\\
&& 4.00 & 27.31 & $3.4872\times 10^{-3}$ & 1.4289 & 0.4288 & 66.86 &5.01\\    \hline  \hline  
\end{tabular}
\tablefoot{$\varepsilon$ is the accretion efficiency of the NS; (a) initial mass of the MS star; (b) initial period of the systems; (c) final period of the systems; (d) total hydrogen mass; (e) final mass of the NS; (f) mass of sdBs; (g) RV semi-amplitude of systems; (h) spin period of the NS.}
\end{table*}

\begin{table*}
\caption{\label{t7} Evolutionary consequence for a binary of a $3.2 \,\mathrm{M_\odot}$ star + $1.25 \,\mathrm{M_\odot}$ NS with $\varepsilon = 0.9$.}
\centering
\begin{tabular}{ccccccccc}
 \hline \hline
$M_\mathrm{NS}^{i}(\mathrm{M}_\odot)$ &$M_\mathrm{d}^{\mathrm{i}}(\mathrm{M}_\odot)^{a}$& $P_\mathrm{i}(\mathrm{d})^b$ & $P_\mathrm{f}(\mathrm{d})^c$ & $M_\mathrm{H}(\mathrm{M}_\odot)^d$ & $M_\mathrm{NS}(\mathrm{M}_\odot)^e$ & $M_\mathrm{sdB}(\mathrm{M}_\odot)^f$ & $K(\mathrm{km\ s^{-1}})^g$ &$P_{\mathrm{NS}}(\mathrm{ms})^h$\\
\hline
&& 1.50 & 8.67 & $0.6002\times 10^{-3}$ & 1.6497 & 0.3257 & 108.61 &0.72 \\
&& 1.60 & 8.10 & $1.8289\times 10^{-3}$ & 1.3226 & 0.3740 & 98.58 &2.46\\
&& 1.70 & 8.17 & $1.9870\times 10^{-3}$ & 1.3140 & 0.3829 & 97.65 &2.70\\
1.25 &3.2& 2.00 & 8.67 & $2.4522\times 10^{-3}$ & 1.3001 & 0.4004 & 94.59 &3.24 \\
&& 3.00 & 11.69 & $3.0752\times 10^{-3}$ & 1.2857 & 0.4193 & 84.52 &4.16\\
&& 4.00 & 15.00 & $3.3815\times 10^{-3}$ & 1.2803 & 0.4264 & 77.40 &4.71\\
&& 5.00 & 18.40 & $3.5647\times 10^{-3}$ & 1.2774 & 0.4300 & 72.12 &5.07\\   \hline  \hline  
\end{tabular}
\tablefoot{$M_\mathrm{NS}^{i}$ is the initial mass of the NS; (a) initial mass of the MS star; (b) initial period of the systems; (c) final period of the systems; (d) total hydrogen mass; (e) final mass of the NS; (f) mass of sdBs; (g) RV semi-amplitude of systems; (h) spin period of the NS.}
\end{table*}

\begin{table*}
\caption{\label{t7}Consequences from the common envelope channel}
\centering
\begin{tabular}{cccccccc}
 \hline \hline
$M_\mathrm{d}^{\mathrm{i}}(\mathrm{M}_\odot)^a$&$P_\mathrm{i}(\mathrm{d})^b$ & $P_\mathrm{f}(\mathrm{min})^b$  & $M_\mathrm{sdB}(\mathrm{M}_\odot)^d$ & $K(\mathrm{km\ s^{-1}})^e$  &$f_\mathrm{0}(\mathrm{Hz})^{f}$&$h^{'}_\mathrm{Gw}(\mathrm{Hz^{-1/2}})^{g}$ & $t(\mathrm{Myr})^h$\\
\hline
&10                        & 25.92                 & 0.3599                  & 778.90    & $1.27887\times 10^{-3}$   &  $  1.48151\times 10^{-18}$      &$2.5938\times 10^{0}$        \\
 3.2 &20                          & 59.04                 & 0.3768                & 589.54  & $5.65211\times 10^{-4}$     &  $   1.17768\times 10^{-18}$    &$2.1931\times 10^{1}$               \\
&30                          & 96.48                  &0.3889               & 497.50   & $ 3.44314\times 10^{-4}$    &  $  1.02808\times 10^{-18}$     &$7.9857\times 10^{1}$             \\
\hline

&12                       & 20.16               & 0.4620                  & 822.36      &  $1.68480\times 10^{-3}$   &  $  2.04599\times 10^{-18}$   &$9.8716\times 10^{-1}$    \\
4.0& 20                          & 36.00                  & 0.4675                & 674.97   &  $9.37043\times 10^{-4}$     &  $ 1.70084\times 10^{-18}$    &$4.6680\times 10^{0}$       \\
& 35                         & 67.68                 & 0.4836                  & 539.78    &  $ 4.87594\times 10^{-4}$    &  $  1.41133\times 10^{-18}$     &$2.5831\times 10^{1}$     \\
& 50                         & 109.44                    & 0.4974                   & 460.13 &  $3.06448\times 10^{-4}$   &  $ 1.24030\times 10^{-18}$         &$8.6871\times 10^{1}$     \\

\hline
& 30                        & 33.12                  & 0.6049                     & 659.34     &  $1.00674\times 10^{-3}$    &  $ 2.20153\times 10^{-18}$        &$3.0505\times 10^{0}$      \\
&40                          & 43.20                   & 0.6095                   & 599.54    &  $7.60428\times 10^{-4}$  &  $2.01874\times 10^{-18}$           &$6.4026\times 10^{0}$        \\
5.0 &50                          & 56.16                  & 0.6169                     & 550.29    &  $ 5.92342\times 10^{-4}$   &  $ 1.87770\times 10^{-18}$         &$1.2329\times 10^{1}$        \\
&60                          & 70.56                      & 0.6242                      & 509.90  &  $4.74609\times 10^{-4}$  &  $ 1.76230\times 10^{-18}$          &$2.2031\times 10^{1}$          \\
& 70                         & 86.40                       & 0.6316                     & 474.82  &  $3.86063\times 10^{-4}$   &  $1.66258\times 10^{-18}$         &$3.7808\times 10^{1}$           \\
& 80                         & 103.68                      & 0.6385                     & 445.77    &  $ 3.21636\times 10^{-4}$  &  $1.57977\times 10^{-18}$         &$6.0924\times 10^{1}$        \\
  \hline  \hline  
\end{tabular}
\tablefoot{(a) Initial mass of the MS star; (b) initial period of systems; (c) final period of systems; (d) mass of sdBs; (e) RV semi-amplitude of systems; (f) frequency of the GW; (g) root spectral density of the GW strain; (h) timescale of the merger.}
\end{table*}

\bibliographystyle{aa}
\bibliography{My_Collection}
\end{document}